\documentclass[prd,twocolumn,showpacs,nofootinbib, aps,10pt,superscriptaddress,amsmath,amssymb]{revtex4-1}

\usepackage{graphicx}
\usepackage{dcolumn}
\usepackage{bm}
\usepackage{epstopdf}
\usepackage{mathtools}
\usepackage{natbib}
\usepackage{captcont}
\usepackage{booktabs}
\usepackage{threeparttable}

\begin{document}
\title{Damping of gravitational waves in a viscous Universe and its implication for dark matter self-interactions}
\author{Bo-Qiang Lu}\email{bqlu@itp.ac.cn}
\affiliation{State Key Laboratory of Theoretical Physics,
Institute of Theoretical Physics, Chinese Academy of Sciences, Beijing, 100190, P.R. China}
\author{Da Huang}\email{dahuang@fuw.edu.pl}
\affiliation{Institute of Theoretical Physics, Faculty of Physics, University of Warsaw, Pasteura 5, 02-093 Warsaw, Poland}
\author{Yue-Liang Wu}\email{ylwu@itp.ac.cn}
\affiliation{State Key Laboratory of Theoretical Physics, 
Institute of Theoretical Physics, Chinese Academy of Sciences, Beijing, 100190, P.R. China}
\author{Yu-Feng Zhou}\email{yfzhou@itp.ac.cn}
\affiliation{State Key Laboratory of Theoretical Physics, 
Institute of Theoretical Physics, Chinese Academy of Sciences, Beijing, 100190, P.R. China}
\begin{abstract}
It is well known that a gravitational wave (GW) experiences the damping effect when it propagates in a fluid with nonzero shear viscosity. In this paper, we propose a new method to constrain the GW damping rate and thus the fluid shear viscosity. By defining the effective distance which incorporates damping effects, we can transform the GW strain expression in a viscous Universe into the same form as that in a perfect fluid. Therefore, the constraints of the luminosity distances from the observed GW events by LIGO and Virgo can be directly applied to the effective distances in our formalism. We exploit the lognormal likelihoods for the available GW effective distances and a Gaussian likelihood for the luminosity distance inferred from the electromagnetic radiation observation of the binary neutron star merger event GW170817. Our fittings show no obvious damping effects in the current GW data, and the upper limit on the damping rate with the combined data is $6.75 \times 10^{-4}\,{\rm Mpc}^{-1}$ at 95\% confidence level. By assuming that the dark matter self-scatterings are efficient enough for the hydrodynamic description to be valid, we find that a GW event from its source at a luminosity distance $D\gtrsim 10^4\;\rm Mpc$ can be used to put a constraint on the dark matter self-interactions.

\end{abstract}
\maketitle

\section{Introduction}
The existence of gravitational waves (GWs) predicted by Einstein a century ago has recently been confirmed by the detection of the first GW signal GW150914 from a binary black hole (BBH) merger by the Laser Interferometer Gravitational-Wave Observatory (LIGO) \cite{GW150914}, which opens a new era of astronomy and cosmology.
Up to date, six BBH merger GW signals, GW150914, LVT151012 (a lower significance candidate), GW151226, GW170104, GW170608 and GW170814 and one binary neutron star (BNS) merger GW signal GW170817 have been announced by the LIGO and Virgo Collaborations~\cite{GW150914, LVT151012, GW151226, GW170104, GW170608, GW170814, GW170817}. Simultaneous detection of GWs and electromagnetic (EM) radiations from the same source can be used to test the fundamental physics, for instance, the speed of GWs by measuring the arrival delays between photons and GWs over cosmological distances~\cite{Will1998, Nishizawa2016, Li2016}, the equivalence principle by using the Shapiro effect~\cite{Will2014, Kahya2016, Wu2016}, and the Lorentz invariance~\cite{Kostelecky2016, Kostelecky20162017}.
The detection of GWs and their EM counterparts can also tell us about the nature of astrophysical sources~\cite{Woosley2016, Valenti2017, Arcavi2017}. For instance, 
the first joint detection of the GRB 170817A and GW170817 confirmed that a neutron star binary could be the progenitor of a short-duration gamma-ray burst (GRB)~\cite{GW170817APJ}.

In the present study, we will show that the ongoing GW observations can provide us a valuable opportunity to examine how the GWs propagate through the matter and, in turn, to constrain the nature of matter in the Universe \cite{Goswami2017, Baym2017, Weinberg2018, Cai2018}. 
It is well known that in the Friedman-Robertson-Walker metric, GWs propagate freely through a perfect fluid without any absorption and dissipation~\cite{Ehlers1987, Ehlers1996, Weinberg1972}. This is no longer true when the Universe contains some nonideal fluids. As pointed out by Hawking half a century ago~\cite{Hawking1966}, when a nonzero shear viscosity $\eta$ is introduced to the fluid energy-momentum tensor, GWs would be dissipated by matter with a damping rate $\beta \equiv 16\pi G\eta$~\cite{Esposito1971, Madore1973, Prasanna1999}, in which $G$ is the gravitational constant.
Note that the shear viscosity $\eta$ and thus the damping rate $\beta$ vary with the evolution of the Universe due to the change of the matter state. But for the timescale concerned here, we can treat $\eta$ and $\beta$ as constants.
Note also that the bulk viscosity, playing an important role in the evolution of the Universe~\cite{Murphy1973, Arbab1997, Fabris2006, Atreya2017}, is shown not to lead to the GW attenuation~\cite{Prasanna1999, Goswami2017}.

Cosmological and astrophysical observations have shown that about 85\% of matter density in the Universe consists of the cold dark matter (DM)~\cite{Rubin1980, Refregier2003, Massey2007}. More recently, the DM self-interaction (SI) is introduced to explain the small-scale structure problems of the Universe~\cite{Spergel2000}. As shown in Refs.~\cite{Goswami2017, Atreya2017}, if the DM can be treated as a fluid, the DM SI can generate the cosmological shear viscosity. Hence, we can transform the constraint of the GW damping to that on the DM SI cross section.

This work is organized as follows. We first present in Sec.~\ref{s2} the expression of the GW strain in a viscous Universe in which we define the effective distance. In Sec.~\ref{s3} we construct a lognormal likelihood function of effective distances for observed GW events. For the BNS merger GW170817, we also take a Gaussian likelihood for the luminosity distance inferred from the GRB observation. By using the $\chi^2$ statistics, we put constraints on the GW damping rate and the shear viscosity. The constraint on the DM SIs from the GW damping is discussed in Sec.~\ref{s4}. 
Finally, we summarize our conclusions in Sec.~\ref{s5}.

\section{Damping of Gravitational Waves}\label{s2}
The standard cosmology assumes that the Universe is homogeneous and isotropic, so that a GW propagating in such a background can be described by the perturbed Friedmann-Robertson-Walker metric~\cite{Weinberg1972,Landau1962}:
\begin{eqnarray}
ds^2={\rm g}_{\mu\nu}dx^{\mu}dx^{\nu}=-dt^2+a^2(\delta_{ij}+h_{ij})dx^{i}dx^{j},
\end{eqnarray}
where we have assumed a spatially flat Universe. The GW is represented by the transverse and traceless tensor perturbation $h_{ij}$, satisfying $\partial^i h_{ij} = h^i_i = 0$. If the GW moves in the $z$-direction, then the two physical GW degrees of freedom, $h_+$ and $h_\times$, can be given by~\cite{Dodelson2003}
\begin{eqnarray}
h_{ij}= \begin{pmatrix}
h_{+} & h_{\times} & 0\\ 
h_{\times} & -h_{+} & 0\\ 
0 & 0 & 0
\end{pmatrix}\,.
\end{eqnarray}

The energy-momentum tensor of a viscous compressible fluid is given by~\cite{Hawking1966}
\begin{eqnarray}
T_{\mu\nu}=(p+\rho)u_{\mu}u_{\nu}+p{\rm g}_{\mu\nu}-2\eta \sigma _{\mu\nu}-\gamma \theta \kappa _{\mu\nu},
\end{eqnarray}
where $p$, $\rho$, $\eta$ and $\gamma$ denote the fluid pressure, density, shear and bulk viscosities, respectively. $u_{\mu}$ is the fluid four-velocity, $\kappa _{\mu\nu}=g_{\mu\nu}+u_{\mu}u_{\nu}$, and the shear of the fluid is
\begin{eqnarray}
\sigma_{\mu\nu}=\frac{1}{2}[(u_{\mu;\nu}+u_{\nu;\mu})+(u_{\mu}u_{\nu}^{\;\;;k}+u_{\nu}u_{\mu}^{\;\;;k})u_{k}]-\frac{1}{3}\theta \kappa_{\mu\nu}.
\end{eqnarray}
By solving the Einstein's equation $G_{\mu\nu}=8\pi GT_{\mu\nu}$ up to the linear order in $h_{ij}$ we can obtain the expression~\cite{Goswami2017, Esposito1971, Dodelson2003} for the GW strain propagating a luminosity distance $D$ through this viscous fluid as
\begin{eqnarray}\label{Ho}
h_{\alpha}=\frac{A(\omega)}{D}\exp\left(\phi_0+\frac{i\omega D}{a}-\int \frac{i\omega}{a} dt\right )\times e^{-\beta D/2},
\end{eqnarray}
where $\alpha$, $\phi_0$, and $A$ denote the GW polarizations, initial phase, and original amplitude, respectively. 
Note that the damping of the GWs in a viscous fluid is reflected in Eq.~(\ref{Ho}) by the exponential suppression factor $e^{-\beta D/2}$ with $\beta = 16\pi G \eta$ called damping rate, which only depends on the shear viscosity of the fluid, rather than the bulk viscosity or the GW frequency.


Now we define the effective distance
\begin{eqnarray}\label{effDis}
D_{\rm eff}=De^{\beta D/2},
\end{eqnarray}
so that the above GW strain formula in Eq.~(\ref{Ho}) can be rewritten as
\begin{eqnarray}\label{Heff}
h_{\alpha}=\frac{A(\omega(t))}{D_{\rm eff}}\exp\left(i{\phi}'_{0}+\frac{i\omega D_{\rm eff}}{a}-\int \frac{i\omega}{a} dt\right )\,,
\end{eqnarray}
which has exactly the same form as a GW from a source at a effective distance $D_{\rm eff}$ with a new unknown initial phase $\phi_0^\prime \equiv \phi_0+\omega (D-D_{\rm eff})/a$ transmitting in a perfect fluid without damping. Note that the observed GW source parameters, especially the luminosity distance, released by the LIGO and Virgo Collaborations are based on the standard assumption 
that all the matters are described by perfect fluids. The form of the GW strain formula in Eq.~(\ref{Heff}) indicates that the GW in a viscous Universe might give the same fitting results, but with the luminosity distance $D$ replaced by the effective distance $D_{\rm eff}$. In light of this insight, the information of the luminosity distance for each GW event published by LIGO and Virgo can be directly applied to the corresponding effective distance in a viscous Universe, which can be further used to constrain the GW damping rate $\beta$ and the fluid shear viscosity $\eta$. 


\section{Data analysis}\label{s3}
\begin{figure}[ht]
\includegraphics[width=70mm,angle=0]{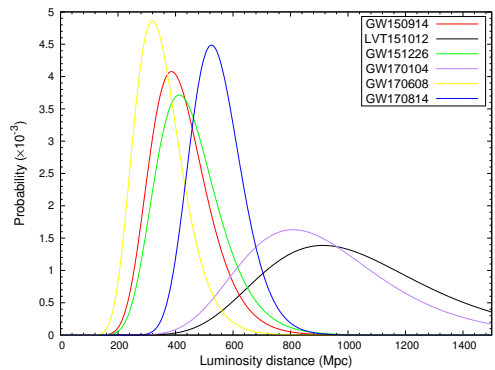}
\caption{Probability distribution of the gravitational wave luminosity distance.}
\label{fig1}
\end{figure}

We notice that the information for each GW source luminosity distance given by LIGO and Virgo Collaborations in Refs.~\cite{GW150914, LVT151012, GW151226, GW170104, GW170608, GW170814, GW170817} follows an approximate lognormal distribution, which has been performed under the assumption that the Universe is filled only perfect fluids. However, in the case with a viscous fluid, these results should be interpreted in terms of the effective distances, as shown in Sec.~\ref{s2}. Hence, the likelihood for the effective distance should be taken as the following lognormal function
\begin{equation}
\begin{aligned}
\mathcal{L}_{{\rm gw},i}\left (D_{{\rm eff},i}|D_{{\rm gw},i},\sigma_{{\rm gw},i} \right )&=\frac{1}{D_{{\rm eff},i}\sigma_{{\rm gw},i}\sqrt{2\pi}}\\
&\times\exp\Bigg (-\frac{\left ( \ln D_{{\rm eff},i}-\ln D_{{\rm gw},i} \right )^2}{2\sigma_{{\rm gw},i}^2} \Bigg )
\end{aligned}
\end{equation}
where $D_{{\rm gw},i}$ and $\sigma_{{\rm gw},i}$ stand for the median values and standard deviations of the measured luminosity distance of this likelihood function for the GW events, which are derived from the data given by LIGO and Virgo in Refs.~\cite{GW150914, LVT151012, GW151226, GW170104, GW170608, GW170814, GW170817} and summarized in Table~\ref{table1}.
We also show the likelihood distributions for the six BBH merger GW events in Fig.~\ref{fig1}.
Note that the free parameters in our fitting include the universal damping rate $\beta$ and the true luminosity distances $D_i$ for GW events, all of which enter the likelihood functions through the effective distances $D_{{\rm eff},i}$. The significance of the GW damping effect can be evaluated with the $\chi^2$ statistics with $\chi^2_i$ for the event $i$ given by
\begin{eqnarray}
\chi^2_i = -2 \ln {\cal L}_{{\rm gw},i} (D_{{\rm eff},i}|D_{{\rm gw},i}, \sigma_{{\rm gw},i}).
\end{eqnarray}
We can obtain the best-fit luminosity distance $D_{{\rm min},i}$ and $\beta$ for each GW event $i$ by minimizing the corresponding $\chi^2_i$. As a result, we find that the minimum $\chi^2_{{\rm min},i}$ are always obtained at $\beta = 0$ with the most favored $D_{{\rm min},i}$ and the values of $\chi^2_{{\rm min},i}$ as given in Table~\ref{table1}. This indicates that there is no evidence for the GW attenuation effects observed so far. The upper limits at 95\% confidence level (CL) on the damping rate $\beta$ can be derived by increasing the $\chi^2_{{\rm min},i}$ by $\Delta \chi^2=3.84$ while fixing the $D_{{\rm min},i}$. We show the upper limits on $\beta$ for observed GW events as in Table~\ref{table1} with the typical constraint of ${\cal O}(10^{-3}\,{\rm Mpc}^{-1})$.

\begin{table}[htbp]
\centering
\caption{Listing of parameters of gravitational wave sources luminosity distance and the fitting results.}
\label{table1}
\begin{threeparttable}
\begin{tabular}{cccccccccccccc}
\toprule
\hline
\hline
GW Event\tnote{\it a} & $D_{\rm gw}$\tnote{\it b} & $\sigma_{\rm gw}$ & $D_{\rm min}$\tnote{\it c} & $\;\;\chi^{2}_{\rm min}$ & $\beta$\tnote{\it d} \\
\hline
\midrule                                                       
GW150914 & $410_{-180}^{+160}$ & 0.246 & 386 & 11 & 2.50 \\
LVT151012 & $1000_{-500}^{+500}$ & 0.301 & 913 & 13 & 1.29 \\
GW151226 & $440_{-190}^{+180}$ & 0.252 & 413 & 11 & 2.39 \\
GW170104 & $880_{-390}^{+450}$ & 0.290 & 809 & 13 & 1.40 \\
GW170608 & $340_{-140}^{+140}$ & 0.249 & 320 & 11 & 3.05 \\
GW170814 & $540_{-210}^{+130}$ & 0.167 & 525 & 11 & 1.25 \\
GW170817 & $40_{-14}^{+8}$ & 0.141 & 39 & 5 & 14.08 \\
\hline
\hline
\bottomrule
\end{tabular}
\begin{tablenotes}
\footnotesize
 \item[\it a] From Refs. \cite{GW150914, LVT151012, GW151226, GW170104, GW170608, GW170814, GW170817}.
 \item[\it b] Median value with 90\% credible intervals of source luminosity distance, in unit of Mpc.
 \item[\it c] Luminosity distance at which $\chi^2$ has minimum value, in unit of Mpc.
 \item[\it d] Upper limit on the damping rate $\beta$ at 95\% CL, in unit of $\rm 10^{-3}\;Mpc^{-1}$.
\end{tablenotes}
\end{threeparttable}
\end{table}

\begin{figure}[ht]
\includegraphics[width=75mm,angle=0]{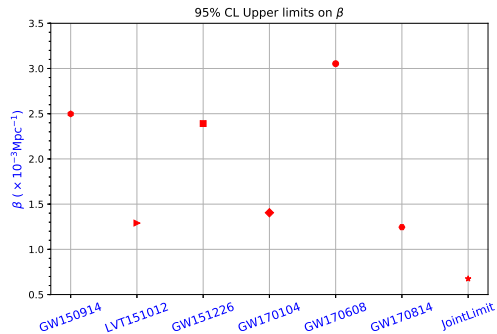}
\caption{Upper limits on the damping rate $\beta$ at 95\% CL for six BBH mergers and their joint analysis.}
\label{fig2}
\end{figure}

The upper limit on the GW damping can be further improved by defining the following joint likelihood function
\begin{eqnarray}
\mathcal{L}_{\rm joint}=\prod _{i}\mathcal{L}_{{\rm gw},i}\left (D_{{\rm eff},i}|D_{{\rm gw},i},\sigma_{{\rm gw},i} \right ),
\end{eqnarray}
and the corresponding $\chi^2_{\rm joint} = -2\ln{\cal L}_{\rm joint}$. The upper limit for the damping rate with $\chi^2_{\rm joint}$ by the same procedure above is $\beta = 6.75\times 10^{-4}\,{\rm Mpc}^{-1}$ at 95\% CL. We also illustrate in Fig.~\ref{fig2} the 95\% CL upper limits on the damping rate $\beta$ for six BBH merger events and the joint analysis.


The advantage of the present method is that we can directly apply the measured luminosity distance informations given by LIGO and Virgo to put constraints on the GW damping rate in a viscous Universe, without requiring a reanalysis of the GW strain data. Moreover, it is interesting to note from Table~\ref{table1} and Fig.~\ref{fig2} that the constraint on $\beta$ becomes more stringent with a lower deviation and a larger observed luminosity distance. As an example, due to its much shorter GW propagating distance, the BNS merger event GW170817 gives the upper bound $\beta \lesssim 14.08\times 10^{-3} \, {\rm Mpc}^{-1}$, which is one order of magnitude less stringent than those from BBH mergers.


The simultaneous detections of the GWs and their EM radiation counterpart open the multi-messenger astronomy era. The recent identification of GW170817 and GRB170817A for the BNS merger event can be used to further strengthen the constraint on the GW damping. In Ref.~\cite{GW170817APJ}, the luminosity distance to the BNS merger host galaxy, NGC~4993, was determined to be $42.9\pm 3.2$~Mpc. Since it is generically assumed that the EM signal is insensitive to the fluid viscosity, we can use this result to directly constrain the true luminosity distance $D$ with the following Gaussian likelihood function:
\begin{eqnarray}
\mathcal{L}_{\rm em}(D|D_{\rm em},\sigma_{\rm em})=\frac{1}{\sqrt{2\pi}\sigma_{\rm em}}\exp\left (-\frac{\left ( D-D_{{\rm em}} \right )^2}{2\sigma_{\rm em}^2} \right )\,,
\end{eqnarray}
where the mean value and deviation are $D_{\rm em}=42.9\rm\;Mpc$ and $\sigma_{\rm em}=3.2\rm\;Mpc$, respectively. With $\chi^2_{\rm tot} = -2(\ln {\cal L}_{\rm gw} + {\cal L}_{\rm em})$ for the BNS merger event which combines both GW and EM informations, we can obtain the minimum value of $\chi^2_{\rm tot,\, min} = 15$ still at $\beta = 0$. However, the 95\% CL upper limit on the GW damping rate $\beta$ is only improved by 1\%, which is too small to be useful due to the short distance of the BNS merger from us.



\section{Implication on Dark Matter Self-Interactions}\label{s4}

Although the collisionless cold DM paradigm can successfully account for the large-scale structure of the Universe~\cite{Bahcall1999}, its predictions via N-body simulations on the small-scale structure seem to conflict with observations on dwarfs~\cite{Oh2011}, low surface brightness (LSB) galaxies \cite{Kuzio2008} and clusters \cite{Newman2013}, known as the cusp-vs-core problem, the missing satellite problem, and the too-big-to-fail problem. One promising solution to all these problems is to introduce the DM SI (For a recent review, please see {\it e.g.} Ref.~\cite{Tulin2017}). 
By assuming that DM self-scatterings are efficient for hydrodynamic description to be valid and assigning a Maxwellian distribution for DM particles, Ref.~\cite{Atreya2017} provided a relation between the DM SI cross section $ \sigma_\chi $ and the shear viscosity $\eta$ 
\begin{eqnarray}
\eta =\frac{1.18\,m_{\chi}\left \langle v \right \rangle}{3 \, \sigma_{\chi}}\,,
\end{eqnarray}
where $m_{\chi}$ and $\langle v \rangle$ are the DM partacle mass and average velocity, respectively. 
We can rewrite the above relation in terms of the GW damping rate $\beta = 16\pi G\eta$ as follows
\begin{eqnarray}
\frac{\sigma_{\chi}}{m_{\chi}}=\frac{6.3\pi G\left \langle v \right \rangle}{\beta}.
\end{eqnarray}
With the typical 95\% upper limits on $\beta \lesssim 10^{-3}\,{\rm Mpc}^{-1}$ obtained in Sec.~\ref{s3}, we find a lower limit on the DM SI to be $\sigma_\chi/m_\chi \sim 10^{-3}\,{\rm cm^2/g}$. However, it has been shown in Ref.~\cite{Atreya2017} that the hydrodynamic description is appropriate at cluster scale only when $\sigma_{\chi}/m_{\chi}\gtrsim 0.1\rm\;cm^{2}/g$, which is not respected by our derived lower limit from the GW damping. Therefore, it is concluded that the current GW measurements cannot give useful bounds on the DM properties, which agrees with the results given in Refs.~\cite{Goswami2017,Baym2017}. 

\begin{figure}[ht]
\includegraphics[width=70mm,angle=0]{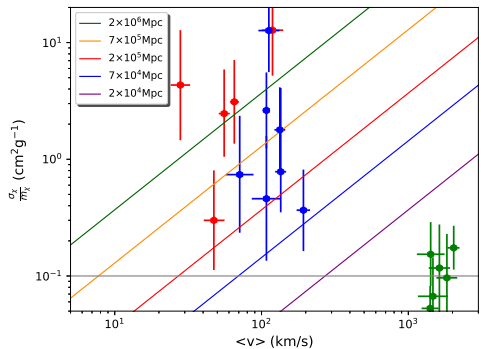}
\caption{Lines from left to right are the potential limits on the $\sigma_{\chi}/m_{\chi}$ from GWs at luminosity distance $2\times 10^{4}\;\rm Mpc$, $7\times 10^{4}\;\rm Mpc$, $2\times 10^{5}\;\rm Mpc$, $7\times 10^{5}\;\rm Mpc$ and $2\times 10^{6}\;\rm Mpc$, respectively. The points represent the positive DM SI signals from dwarf galaxies (red), LSB galaxies (blue), and clusters (green), all of which are obtained from Ref. \cite{Kaplinghat2016}. The horizontal gray line denotes the DM SI condition for the hydrodynamical description to be valid.}
\label{fig3}
\end{figure}
In order to give a sensible constraint, we notice that the GW attenuation becomes strong with the increasing propagation distance, which indicates that the constraints on the GW damping rate can be improved by observing a GW event with a larger luminosity distance. In Fig.~\ref{fig3}, we show the potential bounds on the DM scattering cross section with GWs generated at luminosity distances of $2\times 10^{4}\;\rm Mpc$, $7\times 10^{4}\;\rm Mpc$, $2\times 10^{5}\;\rm Mpc$, $7\times 10^{5}\;\rm Mpc$ and $2\times 10^{6}\;\rm Mpc$, respectively, where the standard deviation $\sigma_{\rm gw}$ in the lognormal likelihood function is assumed to be 0.20. We have also shown in Fig.~\ref{fig3} the DM SI cross sections~\cite{Kaplinghat2016} deduced from the fittings to dwarf galaxies (red), LSB galaxies (blue), and clusters (green), as well as the minimum value of $\sigma_\chi/m_\chi$ for the validity of the DM fluid description (the gray horizontal line). As a result, only when a GW propagating a luminosity distance $D\gtrsim 10^4$~Mpc can the DM SI constraints from the GW damping probe the DM self-scattering solution to the small-scale structure problems.


Let us finish this section by remarking the current status of the GW measurements. In the Advanced LIGO O2 run, the LIGO network has reached the sensitivity for the binary mergers of $10M_{\odot}$ black holes at a distance of 300~Mpc , or those of $30 M_{\odot}$ black holes from 700~Mpc away. 
The Livingston instruments have been sensitive to as far as 100 Mpc for mergers of two $1.4M_{\odot}$ neutron stars.
Moreover, the joint detections by Advanced LIGO and Virgo is promising to improve the ability to localize the GW sources on the sky.
In the near future, the GW reach range will be continuously upgraded by increasing the detector sensitivity over the coming years~\cite{net}.

\section{Summary}\label{s5}

The GWs would be dissipated when propagating through a fluid with shear viscosity.
In the present paper, we propose a new method to study this striking phenomenon by taking an advantage of the ongoing GW observations by LIGO and Virgo.
By defining the GW effective distance encoding the damping rate, we show that the strain formula of a GW moving in a viscous fluid has the same form as that in a perfect fluid, which indicates that we can directly apply the measured GW luminosity distance information to the corresponding effective distance. In the light of this insight,
we construct a lognormal likelihood function of the effective distance from the real data for each GW event, which, together with the $\chi^2$ statistics, makes it possible to constrain the GW damping rate in our Universe.
Consequently, we find that the typical GW damping rate should be bounded by $\beta \lesssim {\cal O} (10^{-3}\;\rm Mpc^{-1})$ at 95\% CL. Our results agree with the previous studies in Ref.~\cite{Goswami2017}, but our method is much simpler since we do not need to reanalyse the raw GW strain data. The method can be further extended to incorporate the constraint on the true GW luminosity distance from 
the EM radiation counterpart of a GW event. This is exemplified by the joint fit to the GW170817 and GRB170817A, both of which were believed to come from the same BNS merger event. However, due to the relatively short distance from the Earth, the final bound on the GW damping is only mildly improved. 

It is expected that the DM SIs can generate the shear viscosity when the DM can be treated with hydrodynamics. By translating the obtained bounds on the GW damping into those on the DM SI cross sections, we find that the current constraints from GWs are typically too weak to be useful. On the other hand, it requires that the luminosity distance of a GW event be as far as ${\cal O}(10^4\,{\rm Mpc})$ in order to give a sensible lower limit on the DM self-scatterings which could potentially solve the cosmological small-scale structure problems.    
  

In addition to the damping due to DM collisions, GWs can also dissipate in the extended gravitational theories, such as the Horndeski theory \cite{Saltas2014, Nishizawa2017} and extra-dimensional theories \cite{Cardoso2003, Deffayet2017}. Thus, our method and results here can also be applied to such theories.


{\it Acknowledgments} BQL, YLW and YFZ are supported by the National Key Research and Development Program of China under Grants No.~2017YFA0402200 and 2017YFA0402204, by the NSFC under Grants No.~11335012 and No.~11475237, and by the Key Research Program of Frontier Sciences, CAS, while DH by the National Science Centre (Poland) research project, decision DEC-2014/15/B/ST2/00108.
%


\end{document}